\documentclass[twocolumn,floatfix,superscriptaddress,a4paper,showkeys,nofootinbib,reprint]{revtex4-1}
\usepackage{epsfig}
\usepackage{latexsym}
\usepackage{xspace}
\usepackage[colorlinks=true,linktocpage=true,linkcolor=blue,citecolor=blue,allcolors=blue]{hyperref}
\usepackage[utf8]{inputenc}
\usepackage{indentfirst}
\usepackage{enumerate}
\usepackage{listings}
\usepackage{color}

\usepackage[caption=false,position=top]{subfig}

\usepackage{amsmath}
\usepackage{amssymb}
\usepackage[english]{babel}
\usepackage{url}

\newcommand{\eq}[1]{\begin{align} #1 \end{align}}
\newcommand{\mean}[1]{\langle #1 \rangle}

\newcommand{\sNN}{\sqrt{s_{\rm NN}}}
\newcommand{\Nacc}{N_{\rm acc}}

\begin{document}

\title{Coordinate versus momentum cuts and effects of collective flow on critical fluctuations}

\author{Volodymyr~A.~Kuznietsov}
    \affiliation{Physics Department, University of Houston, 3507 Cullen Blvd, Houston, TX 77204, USA}
    \affiliation{Department of the high-density energy physics, Bogolyubov Institute for Theoretical Physics, 03680 Kyiv, Ukraine}

\author{Mark I. Gorenstein}
    \affiliation{Department of the high-density energy physics, Bogolyubov Institute for Theoretical Physics, 03680 Kyiv, Ukraine}

\author{Volker Koch}
\affiliation{Nuclear Science Division, Lawrence Berkeley National Laboratory, 1 Cyclotron Road, Berkeley, CA 94720, USA}

\author{Volodymyr~Vovchenko}
\affiliation{Physics Department, University of Houston, 3507 Cullen Blvd, Houston, TX 77204, USA}

\begin{abstract}

We analyze particle number fluctuations in the crossover region near the critical endpoint of a first-order phase transition by utilizing molecular dynamics simulations of the classical Lennard-Jones fluid. We extend our previous study [V.A. Kuznietsov et al., \href{https://doi.org/10.1103/PhysRevC.105.044903}{Phys. Rev. C {\bf 105}, 044903 (2022)}] by incorporating longitudinal collective flow. 
The scaled variance of particle number distribution inside different coordinate and momentum space acceptances is computed through ensemble averaging and found to agree with earlier results obtained using time averaging, validating the ergodic hypothesis for fluctuation observables.
Presence of a sizable collective flow is found to be essential for observing large fluctuations from the critical point in momentum space acceptances.
We discuss our findings in the context of heavy-ion collisions.

\end{abstract}


\maketitle

\section{Introduction}
\label{sec-intro}

Identifying the existence and location of the QCD critical point (CP) at finite baryon density is one of the main goals of the beam energy scans performed with relativistic heavy-ion collisions~\cite{Bzdak:2019pkr}.
Event-by-event fluctuations of the proton number are the primary observable here~\cite{Stephanov:1999zu,Hatta:2003wn}. 
In particular, proton number cumulants are expected to show a non-monotonic collision energy dependence if the QCD critical point exists and heavy-ion collisions are sensitive to it in a narrow collision energy range~\cite{Stephanov:2008qz}.
Experimental measurements performed by the STAR Collaboration within phase I of the RHIC beam energy scan show indications for a non-monotonic collision energy dependence of the proton kurtosis $\kappa \sigma^2 = \kappa_4/\kappa_2$~\cite{STAR:2020tga}, although the experimental error bars are still too large to draw firm conclusions.
On the other hand, 2nd order cumulants of protons were measured with much larger precision~\cite{STAR:2021iop}, and show indications for an excess of the proton number scaled variance at $\sNN \lesssim 20$~GeV relative to baseline expectations due to baryon number conservation and repulsive baryon hard-core~(see \cite{Vovchenko:2023klh} for a recent overview).
Interestingly, measurements at even lower energies, $\sNN = 2.4$~GeV by HADES~\cite{HADES:2020wpc} and $\sNN = 3$~GeV by STAR~\cite{STAR:2021fge} also show indications for the large variance of the proton number distribution,
although these measurements are affected by large volume fluctuation effects unrelated to the CP.
Therefore, the effort to locate the CP with heavy-ion collisions is now mainly focused on collision energies of $\sNN \simeq 2.4-20$~GeV, with future experimental data coming from RHIC BES-II and fixed target programs, as well as the CBM experiment at FAIR~\cite{CBM:2016kpk}.
In addition, several recent effective QCD approaches~\cite{Ratti:2005jh,Ding:2016qdj,Fu:2019hdw,Gunkel:2021oya,Hippert:2023bel,Basar:2023nkp,Goswami:2024jlc} constrained by lattice QCD simulations at $\mu_B = 0$ place the CP into a $T-\mu_B$ range probed by intermediate energy heavy-ion collisions~\cite{Vovchenko:2023klh}.

Interpreting heavy-ion data on event-by-fluctuations is challenging due to many caveats associated with the corresponding measurements.
In particular, direct comparisons of the grand-canonical equilibrium cumulants obtained in most theoretical calculations with experimental measurements are hindered by canonical ensemble effects, the difference between coordinate and momentum space cuts, non-equilibrium dynamics and finite-size effects, and other caveats.
Therefore, a dynamical description of critical fluctuations is required to make meaningful conclusions based on experimental data.
A dedicated effort is underway to incorporate critical fluctuations into relativistic hydrodynamics~\cite{Stephanov:2017ghc,Pradeep:2022mkf}, as well as hadronic transport with mean fields~\cite{Sorensen:2020ygf} or molecular dynamics with a critical point~\cite{Kuznietsov:2022pcn}.
There are also separate studies on the impact of the first-order phase transition on fluctuations~\cite{Steinheimer:2012gc,Savchuk:2022msa,Kuznietsov:2023iyu}, as well as the production of clusters~\cite{Sun:2017xrx,Shuryak:2020yrs}.

\begin{figure*}[t]
    \includegraphics[width=.30\textwidth]{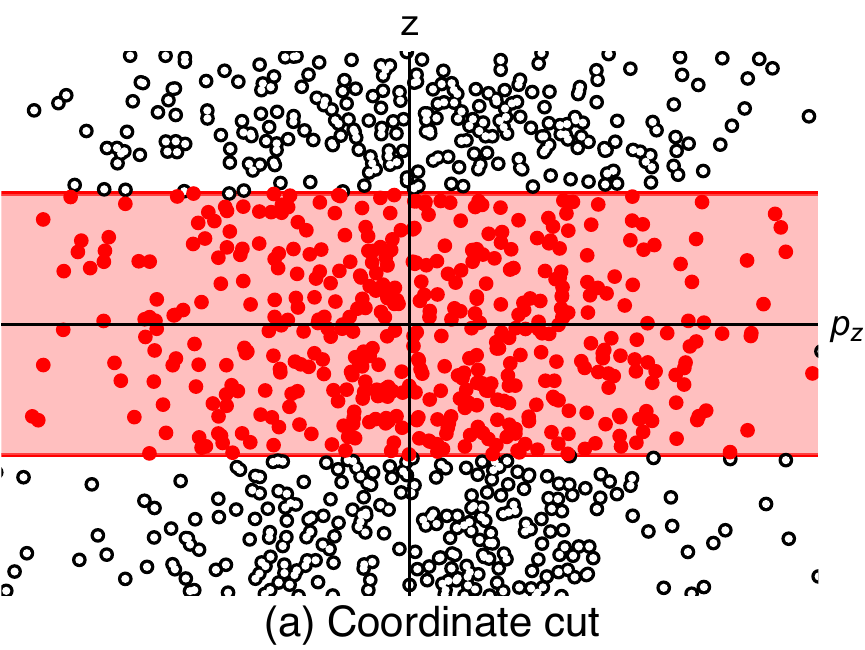}
    \hskip5pt
    \includegraphics[width=.30\textwidth]{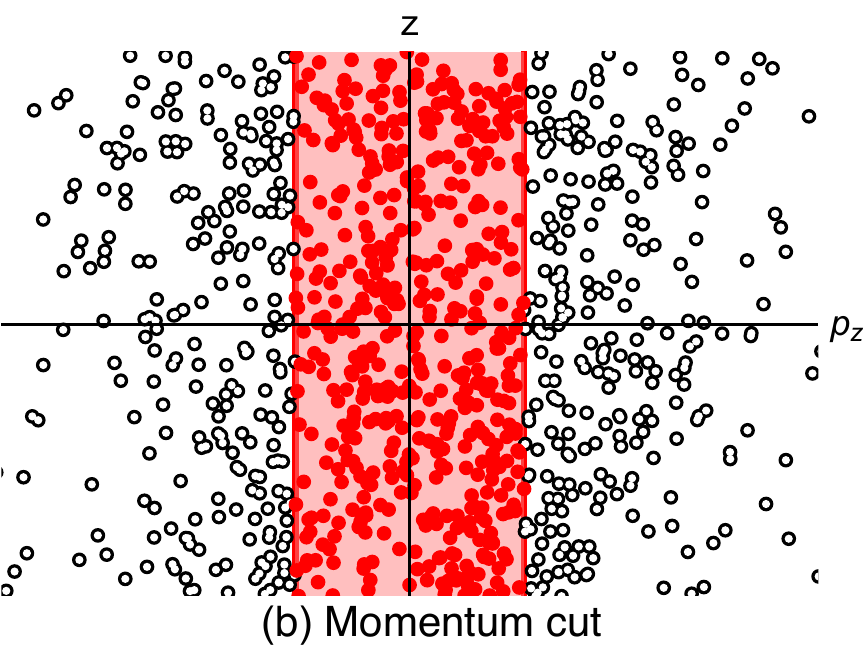}
    \hskip5pt
    \includegraphics[width=.30\textwidth]{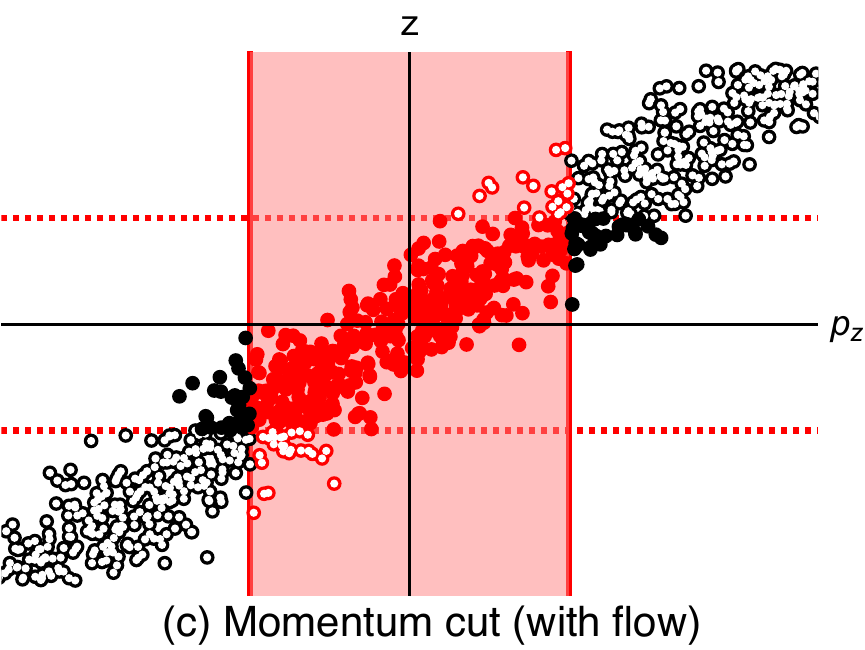}
    \caption{Red points depict the particles inside longitudinal coordinate (a) and momentum space acceptance without (b) and with (c) longitudinal collective flow.
    }
    \label{fig-cuts}
\end{figure*}

In previous work~\cite{Kuznietsov:2022pcn}, we used molecular dynamics (MD) simulations of the Lennard-Jones (LJ) fluid to study the behavior of particle number fluctuations near a CP from the 3D-Ising universality class in a microscopic setup.
The simulations were performed on the crossover side of the transition and confirmed the large imprint of the CP in the variance of particle number inside a coordinate space subsystem. 
However, the large fluctuations of the particle number were completely washed out after coordinate cuts were replaced by momentum cuts~(Fig.~\ref{fig-cuts}).
Since the simulations were performed in a uniform periodic box, there were no correlations between the particles' coordinates and momenta, hence the loss of the CP signal in momentum space.

In the present work, we extend our previous study to conditions appropriate for heavy-ion collisions.
First, we replace time averaging with ensemble averaging by simulating many events with random initial conditions.
In this way, we verify whether the ergodic hypothesis extends to critical fluctuations~\cite{Landau1996}. 
This question is particularly relevant in the context for heavy-ion collisions, where fluctuations are studied on event-by-event basis and hence correspond to ensemble averaging.
Second, we incorporate longitudinal flow, which correlates longitudinal momenta~(rapidities) and coordinates~(space-time rapidities) of particles~(as depicted in panel (c) in Fig.~\ref{fig-cuts}. 
This allows us to establish whether large fluctuations survive in typical rapidity acceptances realized in heavy-ion measurements.

The paper is organized as follows. In Sec.~\ref{sec-setup} we briefly describe the molecular dynamics framework and the simulation setup. 
In Sec.~\ref{sec-coord}, we present the results for coordinate space fluctuations and verify the ergodic hypothesis.
In Sec.~\ref{sec-momentum}, we introduce longitudinal flow and study the behavior of fluctuations at different collision energies and rapidity acceptances. 
We also explore the effect of critical slowing down and compare our results to experimental data.
We summarize our findings in Sec.~\ref{sec-discussion}.

\section{Simulation setup}
\label{sec-setup}

\subsection{Lennard-Jones fluid}

The LJ fluid corresponds to a system of classical non-relativistic particles interacting via the following potential
\eq{
V_{\rm LJ}(r) = 4\varepsilon\left[\left(\frac{\sigma}{r}\right)^{12} - \left(\frac{\sigma}{r}\right)^{6}\right].
}
Here the first term corresponds to the repulsion at short distances while the second term models intermediate range attraction.
The two parameters -- $\sigma$ and $\varepsilon$ -- correspond to the size of the repulsive core and the depth of the attractive well, respectively, and define the corresponding length and energy scales in the system.

It is customary to use dimensionless variables by defining the reduced temperature $\tilde T = T / (k_B \varepsilon)$ and density $\tilde n = n \sigma^3$.
The particle mass defines the dimensionless time variable, $\tilde t = t \sqrt{\varepsilon/(m\sigma^2)}$.
Most properties of the LJ fluid, including the phase diagram in temperature/density plane, become independent of $\sigma$ and $\varepsilon$ in these variables.

Although the equation of state of LJ is not known exactly, it has been studied extensively with molecular dynamics simulations.
The phase diagram of the LJ fluid contains a rich phase structure, including a first-order liquid-gas phase transition with a CP in 3D-Ising universality class~\cite{STEPHAN2020112772}, located at $\tilde T_c = 1.321 \pm 0.007$ and $\tilde n_c = 0.316 \pm 0.005$~\cite{doi:10.1021/acs.jcim.9b00620}.

\subsection{Molecular dynamics}

MD simulations proceed by numerically integrating Newton's equations of motion. The simulations are performed using the Velocity-Verlet integration method for the system of $N$ particles with periodic boundary conditions in the minimum-image convention form\footnote{One can see details of method in \cite{Allen2017} and find the simulation setup source in \cite{LJgithub}}.
In the previous work~\cite{Kuznietsov:2022pcn}, we used the simulations to study the behavior of particle number fluctuations along the supercritical isotherm $\tilde T = 1.06 \tilde T_c$.
This was achieved by performing the simulations for a long period of time at each value of particle number densities and computing the moments of particle number distribution as time average.

In the present work, we explore the same conditions of temperature and density as in Ref.~\cite{Kuznietsov:2022pcn} and use the same GPU-accelerated MD simulation code from~\cite{LJgithub}.
We refer to Sec. III of Ref.~\cite{Kuznietsov:2022pcn} for the details of MD simulation framework.
The key difference to Ref.~\cite{Kuznietsov:2022pcn} is that here we calculate the observables as ensemble averages, namely, by performing a large number of MD simulations at each density, each simulation initialized with random initial conditions.
In this way we are able to compare ensemble averaging with time averaging in Ref.~\cite{Kuznietsov:2022pcn} and study equilibration dynamics at different conditions of particle number density. 
Our simulations here are performed for $N = 400$, which approximately corresponds to the total number of baryons in central collisions of heavy ions when the production of baryon-antibaryon pairs is negligible.

\subsection{Workflow}

\subsubsection{External conditions}

We perform simulations at three points in the phase diagram. They all correspond to the same temperature of $\tilde T = 1.4 \simeq 1.06 \, \tilde T_c$ but different values of the number density: (i) $\tilde n = 0.1 \simeq 0.32 \, \tilde n_c$~(dilute), (ii) $\tilde n = 0.3 \simeq 0.95 \, \tilde n_c$~(critical), and (iii) $\tilde n = 0.6 \simeq 1.90 \, \tilde n_c$~(dense).
The value of the density determines the length of the simulation box, $\tilde L = (N / \tilde n)^{1/3}$, where $N = 400$.
The simulations are performed in the microcanonical ensemble, where the energy per particle $\tilde u = \tilde U / N$, rather than the temperature $\tilde T$ is a fixed quantity strictly conserved throughout the evolution.
To achieve the desired mapping of the microcanonical simulation to the desired $(\tilde T, \tilde n)$ point on the phase diagram, we initialize the system with the energy per particle $\tilde u$ that matches the value from the LJ
equation of state~(see Ref.~\cite{Kuznietsov:2022pcn} for the details on this mapping).
We cross-check that the average value of the kinetic temperature during the simulation matches $\tilde T = 1.4$ to a relative accuracy of about 1\% once equilibrium is reached.

\subsubsection{Initial conditions}

For each $(\tilde T, \tilde n)$ point, we perform approximately 32000 simulations with random initial conditions.
The sampling of initial conditions proceeds as follows:
\begin{enumerate}
    \item The coordinates of all $N$ particles are sampled uniformly within the simulation box of length $\tilde L$.
    Whenever we sample the coordinates of a particle, we check its overlap with any of the previously sampled particles by requiring that the distance to any other particle is larger than 0.9$\sigma$. If an overlap is detected, the coordinates of this particle are rejected and re-sampled until there are no overlaps.
    This step is necessary to maintain stability in the initial state by avoiding large potential energy due to the overlap of any two particles.

    \item The momenta of the particles are sampled independently for each particle from the Maxwell-Bolztmann distribution corresponding to the temperature of $\tilde T$.

    \item For each spatial direction, the momentum components of each particle are shifted by a constant amount such that the total momentum in the system is zero.

    \item The momenta of each particle are rescaled by a constant factor such that the total energy of the system matches the desired input value of $\tilde U$.
    
\end{enumerate}

\subsubsection{MD simulation}

Each event is propagated from the initial time $\tilde t = 0$ to $\tilde t = 100$ by solving the equations of motion with the GPU-accelerated MD solver~\cite{LJgithub}.
We use a time step size of $\Delta \tilde t = 0.004$ for $\tilde n = 0.32\tilde n_{\rm c}$ and $0.95\tilde n_{\rm c}$, and a smaller value of $\Delta \tilde t = 0.002$ for $\tilde n = 1.9\tilde n_{\rm c}$. These values were found to be sufficient to maintain the numerical stability and accuracy of the simulations, which we verified by monitoring the conservation of energy $\tilde U$ throughout the simulation.

The coordinates and momenta of all particles in each event are written to file with a time step of $\Delta t_{\rm out} = 1$ for further processing and analysis.

\subsubsection{Analysis}

The files with the events are processed to analyze the behavior of particle number fluctuations in various setups. 
This is achieved by computing the particle numbers $\Nacc$ in the desired acceptances in each event, then computing
the corresponding scaled variance
\eq{\label{eq:wtil}
\tilde w[\Nacc] = \frac{1}{1-\alpha} \times \frac{\mean{\Nacc^2} - \mean{\Nacc}^2}{\mean{\Nacc}}
}
from the sample. Here $\alpha = \mean{\Nacc} / N$ is the fraction of the whole system inside the acceptance and $\frac{1}{1-\alpha}$ is the correction factor due to global baryon number conservation, as derived in Ref.~\cite{Vovchenko:2020tsr}. 
The moments $\mean{\Nacc}$ and $\mean{\Nacc^2}$ are calculated through
event-by-event averaging\footnote{We estimate their standard error through the Delta theorem by using \texttt{sample-moments} package~\cite{sample-moments}}.

\section{Coordinate space fluctuations and ergodicity}
\label{sec-coord}

\subsection{Ergodicity}

We first look at fluctuations in coordinate space acceptance without any effects of collective flow and expansion.
This is achieved by performing a cut $|\tilde z| < \tilde z_{\rm cut}$ on the longitudinal coordinate of particles.
In this case, the $\alpha$ parameter in Eq.~\eqref{eq:wtil} is known beforehand and is simply equal to the ratio of the subvolume relative to the total volume, $\alpha = 2\,\tilde z_{\rm cut}/\tilde L$.

\begin{figure}[t]
    \includegraphics[width=.49\textwidth]{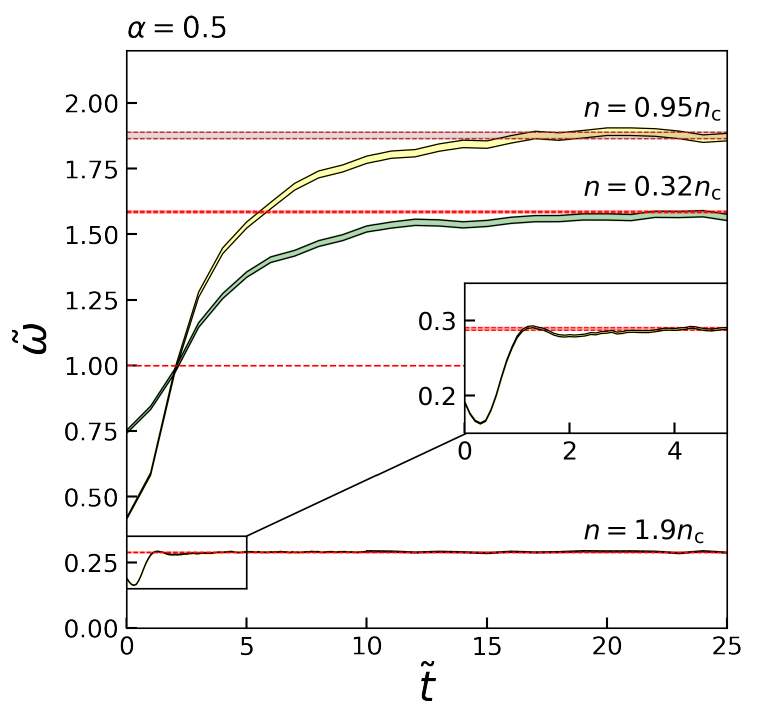}
    \caption{
    Time dependence of the corrected scaled variance $\tilde w_{\alpha}$ of particle number distribution inside longitudinal coordinate space acceptance calculated through ensemble averaging at $T = 1.06 T_c$ for three values of particle number density, $n \simeq 0.32 n_c$~(green band), $n \simeq 0.95 n_c$~(yellow band), and $n \simeq 1.9 n_c$~(brown band). The band width corresponds to the statistical error.
    The horizontal bands correspond to equilibrium expectations from Ref.~\cite{Kuznietsov:2022pcn} computed as time averages.
    The inset zooms into the $\tilde n = 1.9 n_c$ calculation at small times, $\tilde t < 5$.
    }
    \label{fig-ens-average1}
\end{figure}

Figure~\ref{fig-ens-average1} depicts the time evolution of $\tilde w$ for the three densities considered and a fixed value of $\tilde z_{\rm cut}$ corresponding to $\alpha = 0.5$.
In all three cases, one observes saturation of $\tilde w$ values at large times, reflecting the equilibration of fluctuations.
The equilibrium values at large times are consistent within statistical errors with time averages, shown by horizontal bars, from our earlier study~\cite{Kuznietsov:2022pcn}. 
This is true for all values of $0 < \alpha < 1$, not just $\alpha = 0.5$, see Fig.~\ref{fig-wcoord}.
This observation confirms the validity of ergodic hypothesis for particle number fluctuations in pure phases, including the vicinity of the CP.
This confirms the suitability of using event-by-event fluctuations for searching the CP in heavy-ion collisions.

At the initial time, the fluctuations are suppressed relative to the baseline, $\tilde \omega < 1$, with stronger suppression at larger density. 
Recall that the initial conditions correspond to uniform distribution of particles' coordinates, with the additional constraint that no two particles are allowed to overlap.
Prohibiting the overlap corresponds to the effect of hard-core repulsion which suppresses particle number fluctuations~\cite{Vovchenko:2022syc}.

\subsection{Equilibration and critical slowing down}
\label{sec:taueq}

The time it takes for fluctuations to equilibrate is different for different densities, as evident from Fig.~\ref{fig-ens-average1}.
One can characterize the equilibration time more rigorously by considering relaxation time approximation, which is expected to be valid at large times where system is sufficiently close to equilibrium. 
The time dependence of $\tilde w_{\alpha} (\tilde t)$ reads
\eq{\label{eq:tau}
\tilde w_{\alpha} (\tilde t) = \tilde w_{\alpha}^{\rm eq} + \tilde C_{\alpha} e^{-\tilde t / \tilde \tau_\alpha},
}
where $w_{\alpha}^{\rm eq}$ is the equilibrium value, $\tilde \tau_\alpha$ is the equilibration time, and $C_{\alpha}$ is a parameter dependent on initial conditions.

We perform fits to the time dependencies shown in Fig.~\ref{fig-ens-average1} through Eq.~\eqref{eq:wtil} applied to an appropriate time interval where relaxation time approximation is valid. 
We also perform the fit for an additional simulation performed for $n = 0.5 n_c$.
The results are depicted in Table~\ref{tab:taufits}.

The dependence of $\tau_\alpha^{\rm eq}$ on particle number density shows interesting features. It first shows decrease with density, seen by comparing the results for $n = 0.3n_c$ and $n = 0.5n_c$.
Larger relaxation times at lower values of the density can be understood in terms based on the correspondingly large mean free path, $\tau_{\rm mfp} \simeq (\sigma n)^{-1}$.
At large density, $n = 1.9 n_c$, the equilibration time is considerably smaller, $\tilde \tau_\alpha \simeq 0.71$, reflecting fast diffusion in a dense system.
The largest value of
$\tilde \tau_\alpha \simeq 4.06$ is observed near the critical density, $n = 0.95 n_c$, indicating that the density dependence of $\tilde \tau_\alpha$ is a non-monotonic with peak around the critical density.
This observation can be related to the so-called critical slowing down, where it takes a long time for critical fluctuations to reach equilibrium.

One can see that the equilibrium value of $\tilde \omega_{\alpha}^{\rm eq}$ at a density half the critical one, $n = 0.5 n_c$, is almost as large as the one corresponding to $n = 0.95 n_c$.
This begs the question as to why fluctuations at a density considerably below $n_c$ are as large as the fluctuations near the critical point.
This can be explained by stronger finite-size effects at $n = 0.95 n_c$ compared to $n = 0.5 n_c$ when simulations are performed for the same total number of particles.
Indeed, the volume, defined as $V = N / n$, is almost twice larger at $n = 0.5 n_c$.
To verify this assumption we performed additional simulation at $n = 0.5 n_c$ for $N = 210 \simeq 400 \cdot \frac{0.95}{0.5}$, which would make the physical volume at $n = 0.5 n_c$ approximately the same as on for $N = 400$ simulation at $n = 0.95 n_c$.
We find $\tilde \omega_\alpha^{\rm eq} \simeq 1.682$ at $n = 0.5 n_c$ for $N = 210$, which is noticeably below $\tilde \omega_\alpha^{\rm eq} = 1.882$ at $n = 0.95 n_c$.
These results do indicate, however, the challenges associated with controlling the finite-size effects in fluctuations, especially in the presence of the CP.

\begin{table}[t]
\begin{tabular}{c|c|c|c|c}
$n/n_{\rm c}$ & Fit range & $\tilde w_{\alpha}^{\rm eq}$  & $\tilde C_{\alpha}$      & $\tilde \tau_{\alpha}$ \\ 
\hline
\hline
$0.3$                & 
$5 < \tilde t < 25$ &
$1.571 \pm 0.003$ & $-0.823 \pm 0.079$& $3.853 \pm 0.254$ \\ \hline
$0.5$               & 
$5 < \tilde t < 25$ & $1.868 \pm 0.001$ & $-1.361 \pm 0.033$ & $3.584 \pm 0.168$\\ \hline
$0.95$              & 
$5 < \tilde t < 25$ & $1.882 \pm 0.001$ & $-1.132 \pm 0.112$ & $4.055 \pm 0.231$\\ \hline
$1.9$               & 
$2.5 < \tilde t < 10$ & $0.289 \pm 0.001$ & $-0.275 \pm 0.23$ & $ 0.710 \pm 0.160$
\end{tabular}
\caption{
Extracted parameters from the relaxation time approximation~[Eq.~\eqref{eq:tau}] fits to the time dependence of corrected scaled variance $\tilde \omega$ of particle number in coordinate subspace at different densities.
The coordinate space cut corresponds to $\alpha = 0.5$ and the number of particles is $N = 400$ in all cases.
}
\label{tab:taufits}
\end{table}

The equilibration times in Table~\ref{tab:taufits} are given in dimensionless units.
Typical time scales corresponding to hydrodynamic evolution in heavy-ion collisions correspond to 7-10~fm/$c$~\cite{Petersen:2008dd}. Therefore it can be instructive to map the dimensionless LJ units into fm/$c$ to estimate to what extent the large fluctuations can develop in a realistic setup that may be achieved experimentally.
We recall that physical time is related to the dimensionless time as $t = \tilde t \sqrt{(m \sigma^2) / \varepsilon}$.
The value of $\varepsilon$ relevant for heavy-ion collisions can be estimated as $\varepsilon = T_{\rm frz} / \tilde T$ where $T_{\rm frz} \sim 150$~MeV is the typical chemical freeze-out temperature and $\tilde T = 1.4$ is the dimensionless temperature used in the simulations, giving $\varepsilon \sim 107$~MeV.
On the other hand, $\sigma$ corresponds to the hard-core diameter of a nucleon, which we take here to be $\sigma \sim 0.6-0.8$~fm~\cite{Wiringa:1994wb}.
This gives $\sqrt{(m \sigma^2) / \varepsilon} \sim 1.8-2.4$ fm/$c$ as the conversion factor from dimensionless time to fm/$c$ units.
We can thus translate the heavy-ion time scale of $\tau_{\rm HIC} \sim 7-10$~fm/$c$ into dimensionless units: $\tilde \tau_{\rm HIC} \sim 3-5$.
As seen from Figs.~\ref{fig-ens-average1} and~\ref{fig-wcoord}, this is sufficient to fully equilibrate the fluctuations in the dense regime, $n = 1.9 n_c$, where repulsive interactions dominate.
On the other hand, for $\tilde \tau_{\rm HIC} \sim 3-5$ finite-time effects decrease the magnitude in the enhancement of fluctuations at $n = 0.32 n_c$ and $n = 0.95 n_c$ by about a half.

One can note that for the dense system case, $n = 1.9 n_c$, the time dependence of $\tilde \omega_\alpha$ exhibits a non-monotonic oscillation at short initial times, $\tilde t < 2$. 
For this reason, these early times are not included in the fit through Eq.~\eqref{eq:tau}.
To interpret the presence of such oscillation, one can consider the high-density limit, where the coordinates of particles are arranged in a regular array, and where long-range order is present.
In this limit, the motion of the system would correspond to oscillatory perturbations from the equilibrium configuration, and thus make the expected time dependence of $\tilde \omega_\alpha$ to exhibit periodicity and oscillations.
The density $n = 1.9 n_c$ is not yet high enough for the system to be in the crystal phase, but the remnants of the long-range order can cause the initial oscillation of $\tilde \omega_\alpha$.

\subsection{Dependence on acceptance}

\begin{figure*}[t]
    \includegraphics[width=0.93\textwidth]{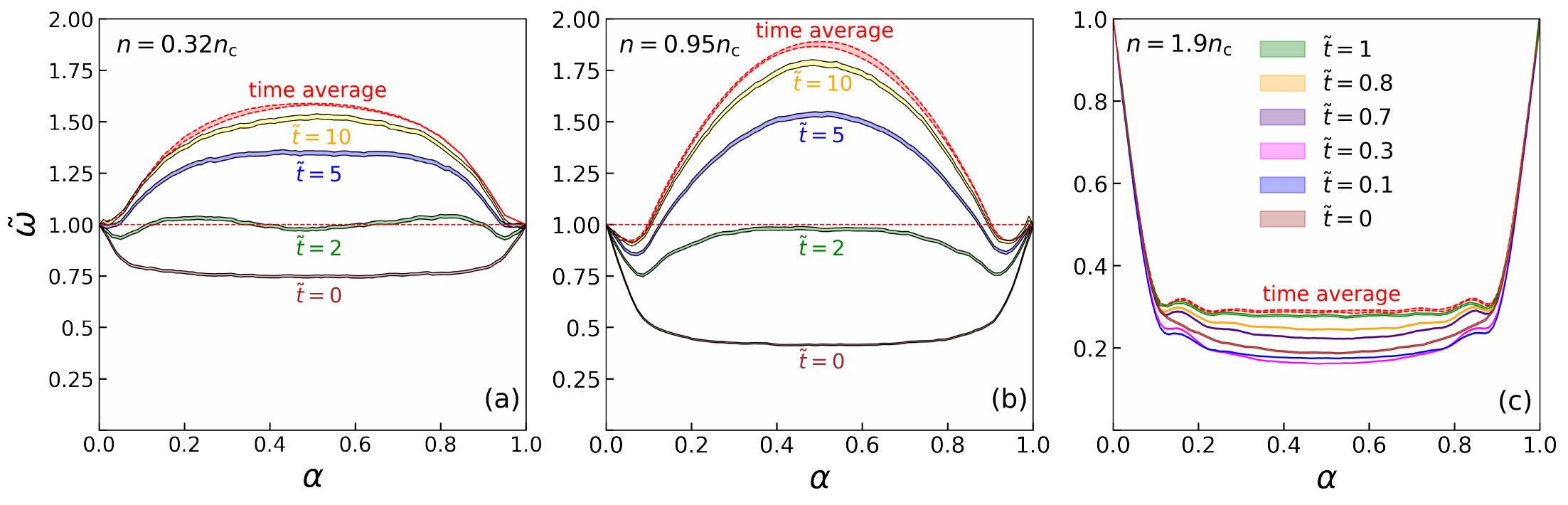}
    \caption{
    The corrected scaled variance $\tilde w_{\alpha}$ of particle number distribution inside longitudinal coordinate space acceptance as a function of acceptance fraction $\alpha$, calculated through ensemble averaging at $T = 1.06 T_c$ for three values of particle number density, $n \simeq 0.32 n_c$~(a), $n \simeq 0.95 n_c$~(b), and $n \simeq 1.9 n_c$~(c).
    Different bands correspond to different values of time after initialization, while 
    the band widths corresponds to statistical error.
    The red bands correspong to equilibrium expectations from Ref.~\cite{Kuznietsov:2022pcn} computed as time averages.
    }
    \label{fig-wcoord}
\end{figure*}

Figure~\ref{fig-wcoord} shows the behavior of $\tilde w_{\alpha}$ as a function of $\alpha$ at different times. Each panel in the figure corresponds to a different value of the density. 
For all values of $\alpha$ and density $n$, one observes that the ensemble average based calculation approaches the time average result of Ref.~\cite{Kuznietsov:2022pcn}.

One can also see that the equilibration time $\tilde \tau_\alpha$ shows some dependence on $\alpha$: at all densities, equilibrium is generally reached faster the further the value of $\alpha$ is from the midpoint value, $\alpha = 0.5$. 
One can also see some non-monotonic behavior of $\tilde w_\alpha$ with respect to $\alpha$ at small values of $\alpha$, for instance at $\simeq 0.1$ (and, by symmetry, at $1 - \alpha \simeq 0.1$) for $n = 0.95 n_c$.
This behavior can be attributed to small longitudinal extent of the coordinate space acceptance that becomes comparable to the size of a single particle. 
Namely, one has
\eq{
\Delta z_{\alpha} = \sigma \alpha \, \tilde L = \sigma \alpha (N/\tilde n)^{1/3}. 
}
For $\tilde n = 0.95 n_c = 0.3$, $N = 400$, and $\alpha = 0.1$ one has $\Delta z_{\alpha} \simeq 1.1 \sigma$, which is comparable to the spatial extent $\sigma$ of a single particle.
A similar effect has been observed in the van der Waals model in Ref.~\cite{Poberezhnyuk:2020ayn} when the system volume becomes comparable to the size of a single particle.

\section{Collective flow and momentum space cuts}
\label{sec-momentum}

\subsection{Incorporating longitudinal flow}

The results from the previous subsection confirm that the presence of a CP leads to large fluctuations of particle number in coordinate space, and the behavior of fluctuations obeys ergodicity.
This confirms that large fluctuation signals of the critical point can be studied both through time and ensemble averaging,
the latter one being particularly relevant to heavy-ion collisions.
However, heavy-ion measurements are performed in momentum space acceptances rather than coordinate space ones.
In a previous work~\cite{Kuznietsov:2022pcn}, we have shown that, in a box calculation, the large fluctuations due to CP point disappear as soon as one replaces coordinate cuts with momentum cuts.
The reason is that particle interactions via the LJ potential occur in the coordinate space, while the momenta and coordinates in a uniform LJ system are uncorrelated. As a result, momentum space fluctuations in LJ system do not show any enhancement due to the CP.
In fact, one only sees an additional suppression, $\tilde w_{\alpha} < 1$, that comes from the global energy-momentum conservation in the microcanonical ensemble.

The situation in heavy-ion collisions is different. 
Due to collective flow, coordinates and momenta of particles at the final stage of hydrodynamic evolution are correlated.
It is thus feasible that large fluctuations can be observed in momentum acceptance due to the presence of such correlation.
Here we introduce the effect of longitudinal flow into our simulations in a simplified way, to evaluate fluctuations in rapidity acceptances typical for heavy-ion collisions.

\begin{figure*}[t]
    \includegraphics[width=.23\textwidth]{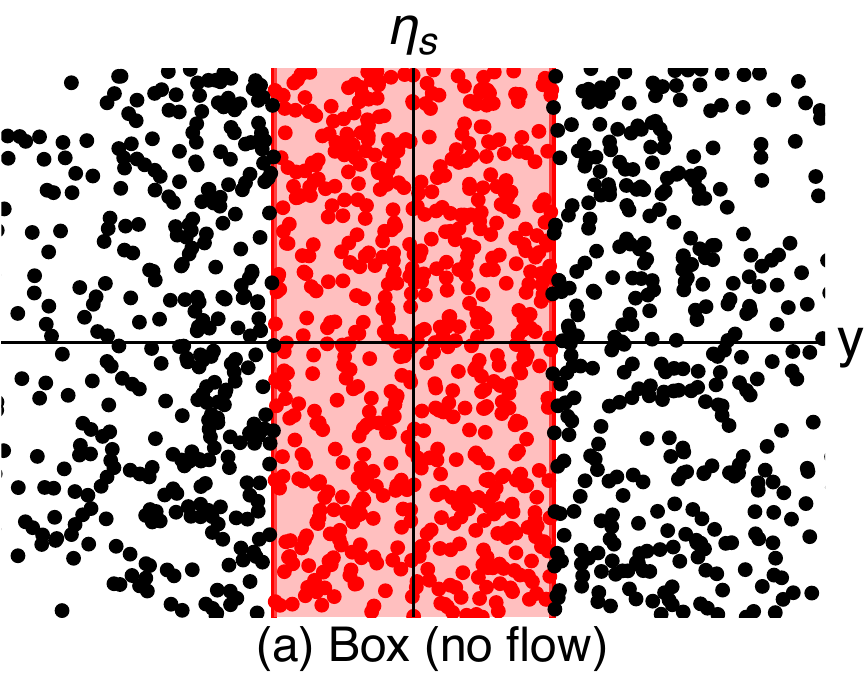}
    \hskip5pt
    \includegraphics[width=.23\textwidth]{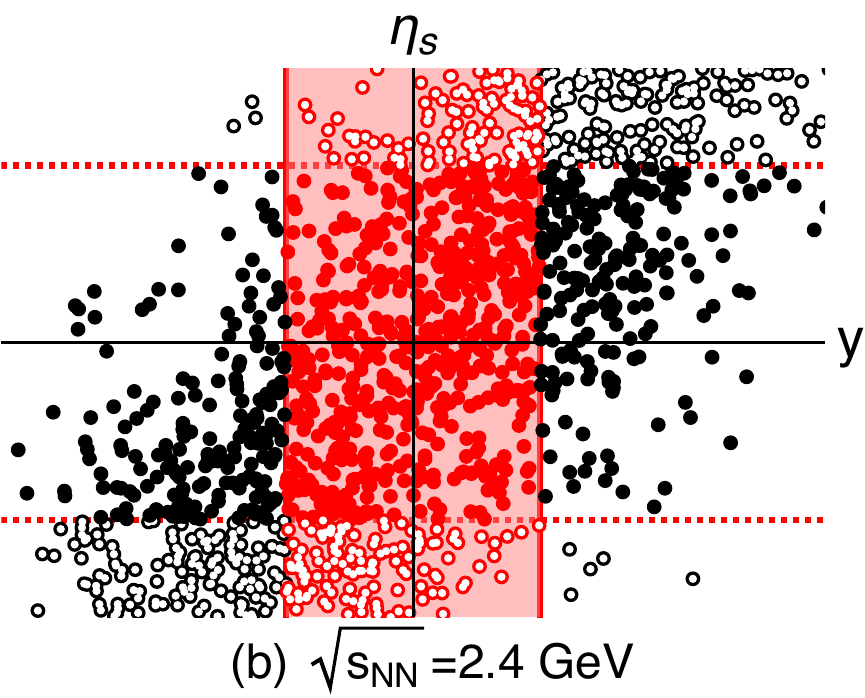}
    \hskip5pt
    \includegraphics[width=.23\textwidth]{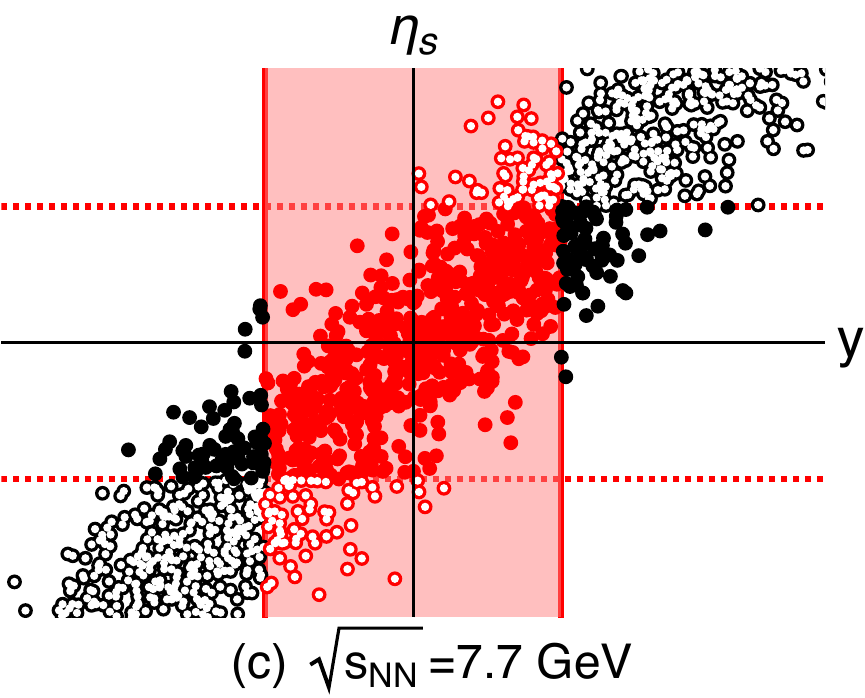}
    \hskip5pt
    \includegraphics[width=.23\textwidth]{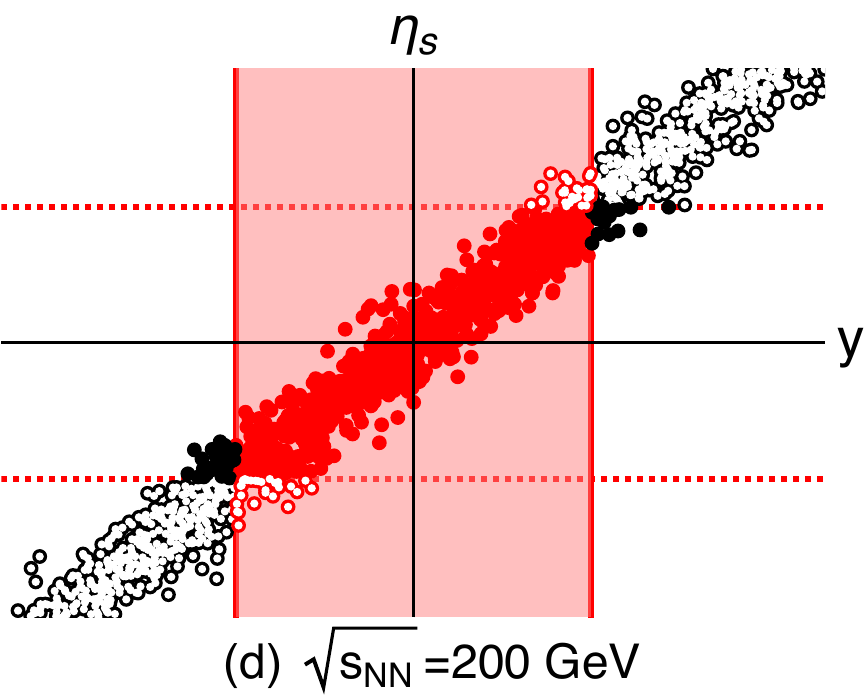}
    \caption{Schematic illustration of the correlation between coordinate space (space-time rapidity $\eta_s$) and momentum space~(longitudinal rapidity $y$) at different collision energies.
    The shaded red region corresponds to momentum space acceptance, $|y| < y_{\rm cut}$, while the dashed horizontal line depict the equivalent coordinate space acceptance.
    The momentum cut is chosen in each case to cover approximately half of all particles, i.e. $\alpha = 0.5$.
    Open red points and solid black points depict, respectively, false positives and false negatives -- the particles that are inside~(outside) momentum acceptance but outside~(inside) the equivalent coordinate acceptance.
    }
    \label{fig-flow}
\end{figure*}

Our considerations are restricted to longitudinal direction only, meaning that transverse momenta of particles are integrated over.
Specifically, we define for each particle the longitudinal rapidity\footnote{The rapidity $y$ and velocity $v_z$ coincide in the non-relativistic limit.} $y$ as a sum collective and thermal components,
\eq{
y = y^{\rm coll} + y^{\rm th}.
}
The LJ fluid simulations in the box setup have no collective motion, thus the velocities $\tilde v_z^{\rm LJ}$ from the simulation define the thermal component of the rapidity, $y^{\rm th} \propto \tilde v_z^{\rm LJ}$.
To define the conversion factor, recall that the $\tilde v_z^{\rm LJ}$ distribution in LJ simulation corresponds to the Maxwell-Bolztmann distribution with a width of $\sigma_{\tilde v_z^{\rm LJ}} = \sqrt{\tilde T}$.
On the other hand, the width of the thermal rapidity distribution in heavy-ion collisions is $\sigma_y = \sqrt{T_{\rm frz} / m_N}$.
Therefore,\footnote{Here we neglected relativistic effects, which allowed us to equate the longitudinal velocity and rapidity thermal components. Such an approximation is justified when $m_N/T_{\rm frz} \gg 1$. For $m_N = 938$~MeV and $T_{\rm frz} = 150$~MeV, one has $m_N/T_{\rm frz} \sim 6$.}
\eq{
y^{\rm th} = \sqrt{\frac{T_{\rm frz}}{m_N \tilde T}} \, \tilde v_z^{\rm LJ}.
}

The collective component of the rapidity is a function of coordinate.
Due to the ultrarelativsitic nature of the motion in the longitudinal direction, it is common to work in Bjorken variable, where instead of the Minkowski time $t$ and the longitudinal coordinate $z$ one uses Milne coordinates, the longitudinal proper time $\tau = \sqrt{t^2 - z^2}$ and the space-time rapidity, $\eta_s = \frac{1}{2} \ln \frac{t-z}{t+z}$, are used instead.
In the Bjorken-like longitudinal flow picture, collective component of the rapidity coincides with the space-time rapidity $y_{\rm coll} = \eta_s$.
To make a connection between the LJ longitudinal coordinate $\tilde z^{\rm LJ}$ and $\eta_s$ at a given collision energy $\sNN$, we make a linear map between the interval $\tilde z^{\rm LJ} \in [-\tilde L / 2, \tilde L / 2]$ and the space-time rapidity extent $\eta_s \in [-y_{\rm cm}^{\rm beam}, y_{\rm cm}^{\rm beam}]$, where
\eq{
y_{\rm cm}^{\rm beam} (\sNN) = \ln \left[ \frac{\sqrt{s_{\rm NN}} + \sqrt{s_{\rm NN} - 4 m_N^2}}{2 m_N} \right],
\label{eq:ycm}
}
is the beam rapidity in the center-of-mass frame of the collision.
Therefore,
\eq{
y^{\rm coll} = \frac{2 y_{\rm cm}^{\rm beam}}{\tilde L} \, \tilde z^{\rm LJ}.
}

Our implementation assumes that the density of particles is flat as a function of space-time rapidity $\eta_s$, i.e. that the system is boost-invariant up to the beam rapidity, $n(\eta_s) \propto \Theta(y_{\rm beam}^{\rm cm} - \eta_s)$. 
In a more involved study, one can explore non-uniform distribution with respect to $\eta_s$, which we leave for future work.

Calculations of fluctuations in longitudinal rapidity acceptance for a given energy, therefore, proceed as follows
\begin{enumerate}
    \item In each event, the rapidity of each particle is calculated through 
    \eq{\label{eq:ytot}
    y = \frac{2 y_{\rm cm}^{\rm beam}}{\tilde L} \, \tilde z^{\rm LJ} + \sqrt{\frac{T_{\rm frz}}{m_N \tilde T}} \, \tilde v_z^{\rm LJ},
    }
    where $y_{\rm cm}^{\rm beam} = y_{\rm cm}^{\rm beam}(\sNN)$ is given by Eq.~\eqref{eq:ycm}.
    We use $T_{\rm frz} = 150$~MeV and $m_N = 938$~MeV/$c^2$.
    \item The number of accepted particles $N_{\rm acc}$ is computed by performing a rapidity cut $|y| < y_{\rm cut}$.
    \item The corrected scaled variance $\tilde w_{\alpha_{y_{\rm cut}}}$ is computed through~\eqref{eq:wtil}, where 
    \eq{
    \alpha_{y_{\rm cut}} = \frac{\mean{N_{\rm acc}}}{N}.
    }
\end{enumerate}

The procedure described above is the simplest one for implementing longitudinal flow into the system, which corresponds to the transformation of a single fireball in a box into an expanding one (see Fig. \ref{fig-flow}).
It relies on the Bjorken picture~[the second term in Eq.~\eqref{eq:ytot}] as well as the absence of event-by-event fluctuations of in the longitudinal flow.
As such, the description must be improved for quantitative applications, especially at lower energies from RHIC beam energy scan.
In the present work, we retain the picture presented above to make a first estimate of the effect of longitudinal flow under the most favorable (and simplified) conditions possible.

\subsection{Fluctuations at fixed $\alpha$}

\begin{figure}[t]
    \includegraphics[width=.45\textwidth]{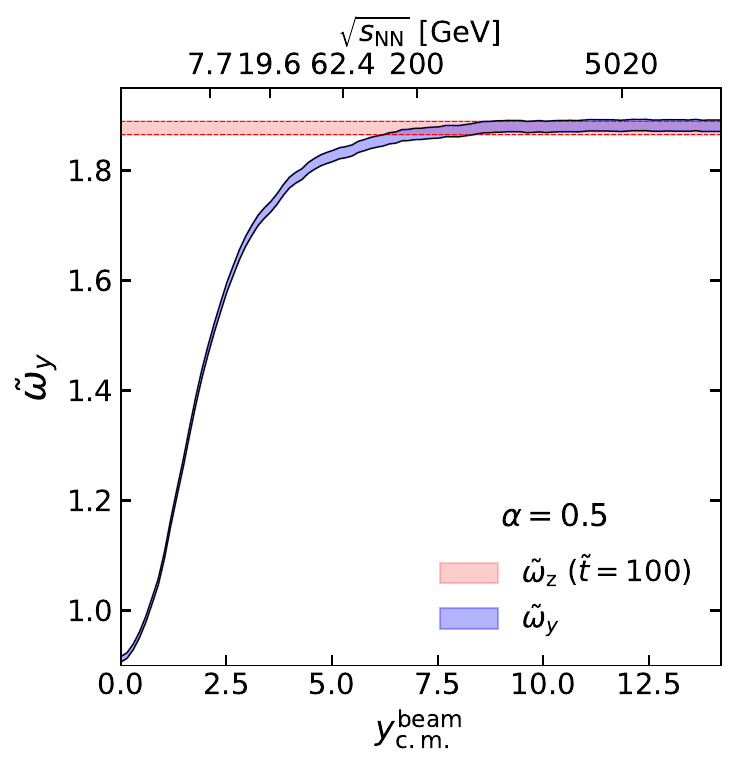}
    \caption{Corrected scaled variance $\tilde \omega_{\alpha}$ of particle number in rapidity acceptance as a function of $y_{\rm cm}$ for a fixed value of $\alpha = 0.5$.
    Calculations are performed for a LJ system of $N = 400$ particle near the CP, $T = 1.06 T_c$ and $n = 0.95 n_c$.
    }
    \label{fig:wybeam}
\end{figure}

We first explore the behavior of fluctuations at different energies for a fixed value of $\alpha$.
We take $\alpha = 0.5$ and vary the value of $y_{\rm cut}$ at each energy to match $\alpha = \mean{N_{\rm acc}} / N = 0.5$.
Figure~\ref{fig:wybeam} shows the resulting dependence of $\tilde w_{\alpha_{y_{\rm cut}}}$ on the beam c.m. rapidity $y_{\rm cm}$ for $n = 0.95 n_c$.
Calculations are performed at large times $\tilde t = 100$ corresponding to an equilibrated system.
$\tilde w_{\alpha_{y_{\rm cut}}}$ monotonically increases with $y_{\rm beam}$ and saturates at a value consistent with the coordinate space result from Sec.~\ref{sec-coord}, shown by the horizontal band.
The result confirms that a strong collective flow allows one to map coordinate space fluctuations to momentum space ones.
Mathematically, this conclusion follows from Eq.~\eqref{eq:ytot}, where the first term becomes dominant at large $y_{\rm cm}$ and thus cuts in rapidity $y$ become equivalent to cuts in coordinate $\tilde z^{\rm LJ}$.
On the other hand, for $y_{\rm beam} \to 0$ we reproduce box simulation results from~\cite{Kuznietsov:2022pcn} where large fluctuations in momentum space are absent.

\begin{figure}[t]
    \includegraphics[width=.44\textwidth]{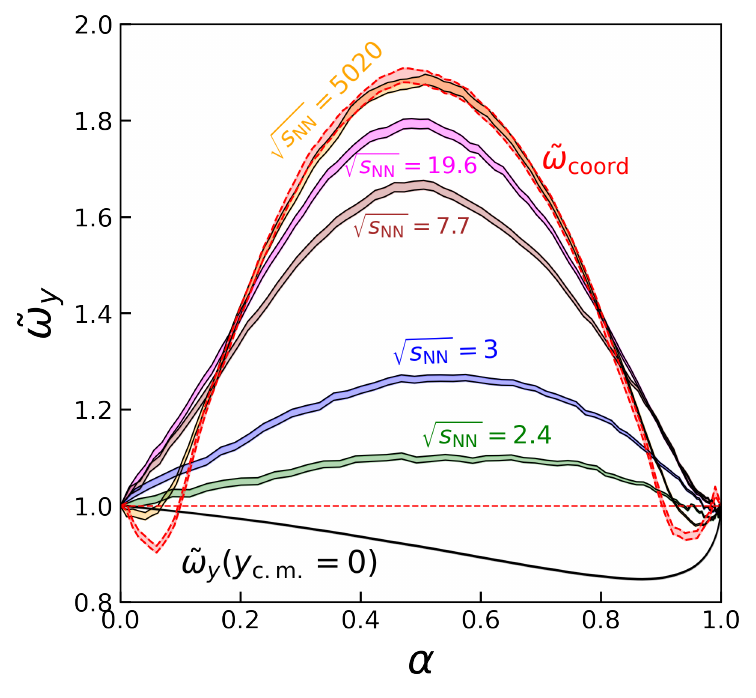}
    \caption{
    Corrected scaled variance $\tilde \omega_{y}$ of particle number in rapidity acceptance as a function of the fixed acceptance fraction $\alpha_y$.
    Calculations are performed for a system of $N = 400$ particles at $T = 1.06 T_c$ and $n = 0.95 n_c$ at large times~($\tilde t = 100)$) where finite-time effects can be neglected.
    Different bands correspond to different collision energies, with the incorporation of collective expansion in the longitudinal direction, as detailed in the text.
    The limiting cases of coordinate [red band, labeled $\tilde \omega_{\rm coord}$] and rapidity acceptance [black line, labeled $\tilde \omega_y (y_{\rm cm} = 0)$] in the absence of collective expansion are also shown.}
   \label{fig:wyalpha}
\end{figure}

The dependence of $\tilde w_{\alpha_{y_{\rm cut}}}$ on $\alpha$ at different energies is shown in Fig.~\ref{fig:wyalpha} and shows a consistent approach toward the coordinate space result at all values of $\alpha_{y_{\rm cut}}$.

\subsection{Fluctuations at fixed $y_{\rm cut}$}

\begin{figure}[t]
    \includegraphics[width=.44\textwidth]{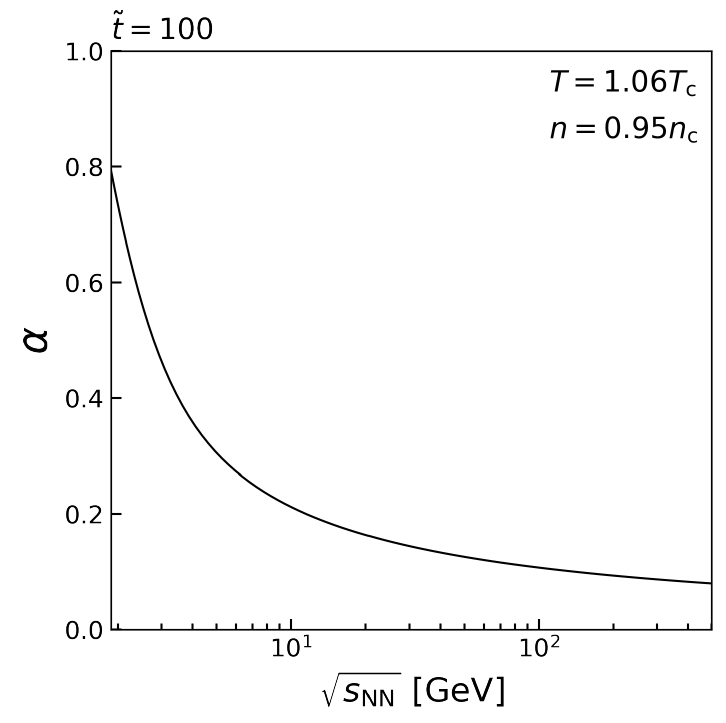}
    \caption{The dependence of acceptance fraction $\alpha = \mean{N_{\rm acc}}/N$ corresponding to the rapidity cut $y_{\rm cut} = 0.5$ on the collision energy, $\sqrt{s_{\rm NN}}$, evaluated for $\tilde t = 100$.}
    \label{fig-alpha_ycut}
\end{figure}

In heavy-ion collisions, the measurements are usually performed in a fixed interval around midrapidity.
For instance, STAR measurements of proton number cumulants~\cite{STAR:2020tga,STAR:2021iop} were done in acceptance $|y| < 0.5$.
Fixing the value of $y_{\rm cut}$ differs from fixing the acceptance fraction $\alpha$: for a fixed value of $y_{\rm cut}$, the value of $\alpha$ will be smaller at larger collision energy, see Fig.~\ref{fig-alpha_ycut}.
This is easy to explain because larger energies lead to a larger total coverage in rapidity. Thus, a fixed rapidity cut covers a smaller fraction of the whole system for larger $\sNN$.

\begin{figure*}[t]
    \includegraphics[width=.9\textwidth]{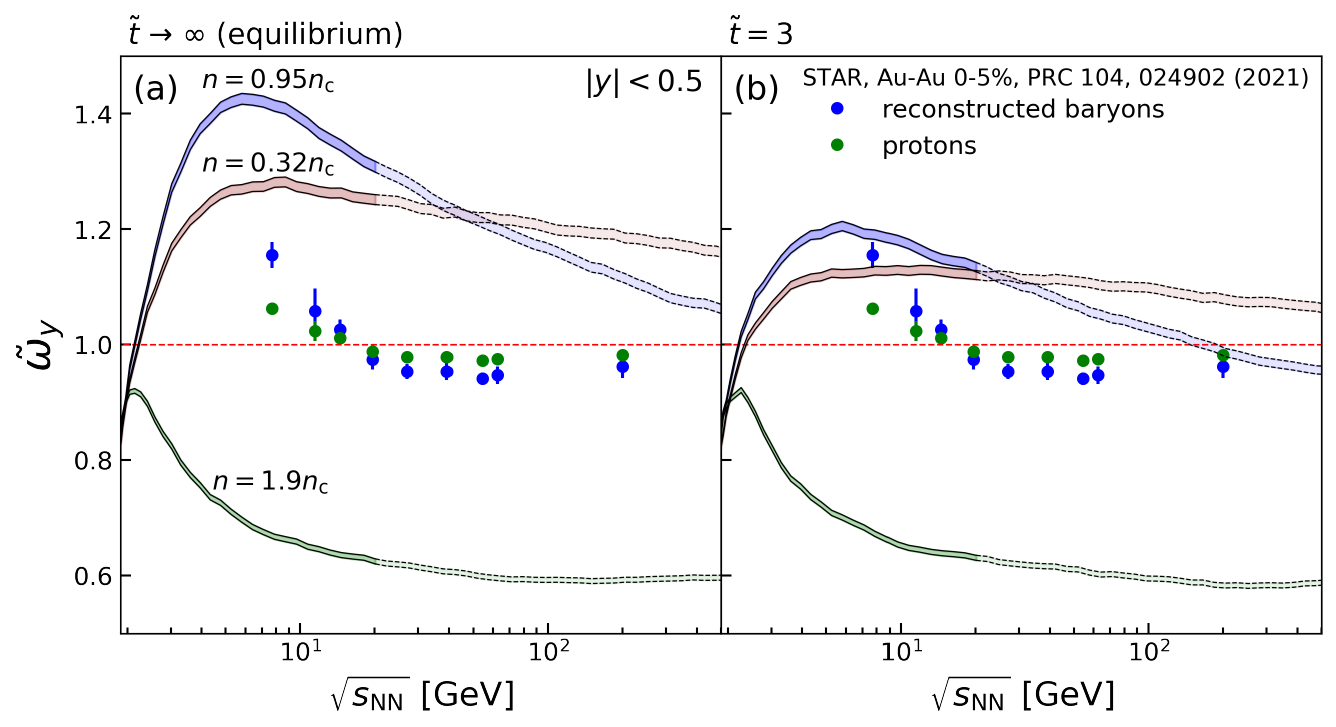}
    \caption{Corrected scaled variance of particle number as a function of $\sqrt{s_{\rm NN}}$ in rapidity acceptance $|y| < 0.5$ calculated through molecular dynamics with collective flow at (a) $\tilde t \to \infty$~(equilibrium) and (b) $\tilde t = 3$~(heavy-ion time scale).
    Different bands have the same meaning as in Fig.~\ref{fig-ens-average1}.
    The bands are shaded at $\sNN > 19.6$~GeV to reflect the absence of antibaryons in our modeling, which is relevant at those collision energies.
    The symbols correspond to the experimental data of the STAR Collaboration~\cite{STAR:2021iop} for protons~(blue) and reconstructed baryon~(red).
    The data are corrected for global baryon conservation by diving over $1-\alpha$ factor, estimated at each collision energy through hydrodynamic simulations from Ref.~\cite{Vovchenko:2021kxx}.
    }
    \label{fig-wsNN}
\end{figure*}

The left panel of Figure~\ref{fig-wsNN} shows the collision energy dependence of nucleon number fluctuations in acceptance $|y| < 0.5$ covering one unit of midrapidity, for the three densities considered. Calculations are performed at the large time, $\tilde t = 100$, corresponding to equilibrium\footnote{Including effects of collective flow and finite system size.} expectations.
Let us focus on the $n = 0.5 n_c$ calculation~(blue band).
This calculation depicts the expected behavior of fluctuations under the assumption that the freeze-out of fluctuations at a given collision energy occurs near the CP. As such, the results should not be considered as predictions of the collision energy dependence measured by the experiment.

Our calculations indicate that the maximum CP signal would be observed at $\sNN \sim 5-7$~GeV, i.e. if the CP is accessible in heavy-ion regime, these collision energies are optimal for observing its signatures in baryon number cumulants based on our description.
This 'sweet spot' in $\sNN$ is an interplay of two effects.
At lower collision energies, the CP signal is diluted due to a weak collective flow and the absence of correlations between coordinates and momenta.
On the other hand, larger collision energies correspond to smaller values of $\alpha$\footnote{In the limit $\sNN \to \infty$ we have $\alpha \to 0$ and thus $\tilde \omega_\alpha \to 1$.}. Thus, the finite-size effects are stronger at large $\sNN$, and these dampen the CP signal~(see Fig.~\ref{fig-wcoord}).
Interestingly, a similar, broader maximum is observed for $n = 0.32 n_c$, where the fluctuations also show enhancement, albeit smaller in magnitude compared to the CP.
At large density, $n = 1.9 n_c$, the fluctuations are suppressed due to repulsive interactions, this suppression decreases monotonically with collision energy.
Of course, given the limitations of our approach (Sec. \ref{Limit}) and the complexity of the system created in heavy-ion collisions at moderate energies, our conclusions could be modified in a more involved approach.
However, we do expect the qualitative interplay between the increase of system size at fixed $y_{\rm cut}$ and the dilution of space-momentum correlations as $\sNN$ is decreased to hold.


It should be noted that our simulations neglect the production of antibaryons, which becomes increasingly relevant at high collision energy.
This can be quantified by the $\bar{p}/p$ ratio, measured by STAR at different collision energies~\cite{STAR:2017sal}. 
Antibaryons can be neglected at $\sNN \lesssim 11.5$~GeV, as STAR has measured
$\bar{p}/p \simeq 0.01$ at $\sNN = 7.7$~GeV and $\bar{p}/p \simeq 0.03$ at $\sNN = 11.5$~GeV. The antiproton fraction becomes more sizable at $\sNN = 19.6$~GeV, where $\bar{p}/p \simeq 0.12$.
For these reasons, our results in Fig.~\ref{fig-wsNN} at energies above $\sNN \geq 19.6$~GeV are shaded to emphasize the absence of antiparticles in our calculations which should not be neglected at these energies. 

We also depict in Fig.~\ref{fig-wsNN} the experimental data of the STAR Collaboration~\cite{STAR:2021iop} on the corrected scaled variance of proton number, $\tilde \omega_p = \omega_p / (1-\alpha_p)$~(green symbols) and reconstructed baryon number, $\tilde \omega_B = \omega_B / (1-\alpha_B)$~(blue symbols)\footnote{Note that here we depict fluctuations of the particle number, rather than the commonly used net particle number.} in the same rapidity acceptance $|y| < 0.5$ as our calculation\footnote{The experimental measurements contain additional cut in transverse momentum, $0.4 < p_T < 2.0$~GeV/$c$. This effect is absent in our calculations.}.
As before, $1-\alpha_{p(B)}$ factors implement the correction due to baryon number conservation.
To correct the data, we take $\alpha_{p(B)}$\footnote{We include contributions of antibaryons when calculating $\alpha_{p(B)}$ for correcting the data.} from Ref.~\cite{Vovchenko:2021kxx} from state-of-the-art (3+1)D hydrodynamic simulations~\cite{Shen:2020jwv}.

We note that baryon fluctuations are not measured directly in the experiment. Instead, we reconstruct $\omega_B$ from the measured $\omega_p$ through the unfolding method from~\cite{Kitazawa:2012at}. We perform this reconstruction to ensure a meaningful correspondence between the measured quantities and those computed in our model. 
The experimental data show enhancement of $\tilde \omega$ with respect to unity at low collision energies and suppression at large energies.
The maximum value of $\tilde \omega_B \simeq 1.15$ is reached at the lowest available BES energy of $\sNN = 7.7$~GeV. 
This indicates enhancement of fluctuations, although the data are considerably closer to the baseline than our equilibrium calculations for $n = 0.32 n_c$ and $n = 0.95 n_c$. 
We would like to emphasize here, however, that our model is not sufficiently sophisticated to draw conclusions from quantitative comparisons with experimental data.
Instead, we make comparisons with data in Fig.~\ref{fig-wsNN} to study qualitative behavior, as well as to estimate the possible magnitude of the CP signal in heavy-ion collisions under the most favorable conditions possible.

At large energies, $\sNN \gtrsim 20$~GeV, the data indicates mild suppression with respect to the baseline, $\tilde \omega_B \simeq 0.95 < 1$. 
This suppression can be attributed to repulsive interactions, which suppress fluctuations~\cite{Gorenstein:2007ep}.
The suppression in the data is not as strong as in our $n = 1.9 n_c$ calculation, where the effects of repulsive interactions are very strong, and where antibaryons are neglected, but was shown in Refs.~\cite{Vovchenko:2021kxx} to be described well by excluded volume effects of moderate strength.

\subsection{Finite-time effects}

Our calculations shown in Fig.~\ref{fig:wyalpha} correspond to large time, $\tilde t = 100$.
Given that equilibration times are much smaller, $\tilde \tau_\alpha \lesssim 4$~(see Table~\ref{tab:taufits}), the calculation essentially corresponds to the equilibrium expectation.
However, the system in heavy-ion collisions is short-lived~($\tau_{\rm HIC} \sim 7-10$fm/$c$), which corresponds to $\tilde\tau_{\rm HIC} \sim 3-5$ in dimensionless units~(see Sec.~\ref{sec:taueq}).
Therefore, it is important to incorporate these finite-time effects, especially for fluctuations near the CP.

\begin{figure}[t]
    \includegraphics[width=.45\textwidth]{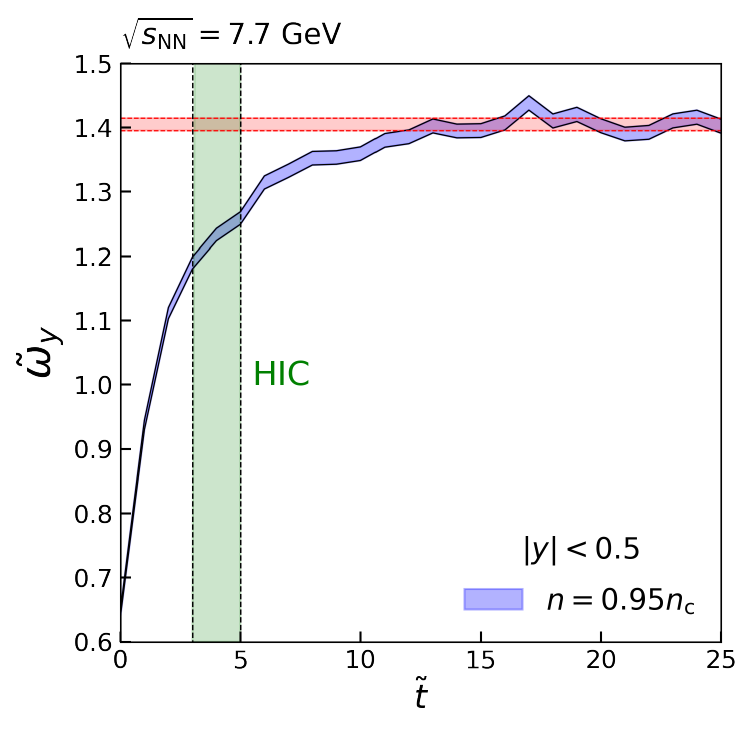}
    \caption{Scaled variance as a function of time for the given $\sqrt{s_{\rm NN}}=7.7$ GeV at fixed volume in the momentum space. The red band represents $\tilde \omega_{y}$ at $\tilde t = 100$, corresponding to the equilibrium expectation where finite-time effects are negligible.
    The vertical green band corresponds to time scales relevant to heavy-ion collisions~(see the text for details).
    }
    \label{fig:wytimedep}
\end{figure}

Here we address this question, focusing on fluctuations in the rapidity space for fixed $y_{\rm cut} = 0.5$ and fixed energy of $\sNN = 7.7$~GeV. Figure~\ref{fig:wytimedep} depicts the time dependence of $\tilde \omega$. This quantity reaches the equilibrium expectation shown by the red band at $\tilde t \gtrsim 10$. At shorter times, however, large deviations from the equilibrium value are seen. In particular, this is the case for $\tilde t \sim 3-5$ relevant for heavy-ion collisions, as discussed above.
Of course, given the limitations of our approach, in particular the difference between initial conditions in our simulations and those in heavy-ion collisions, these estimates can give only a qualitative picture of the possible finite-time effects.

\begin{figure}[t]
    \includegraphics[width=.45\textwidth]{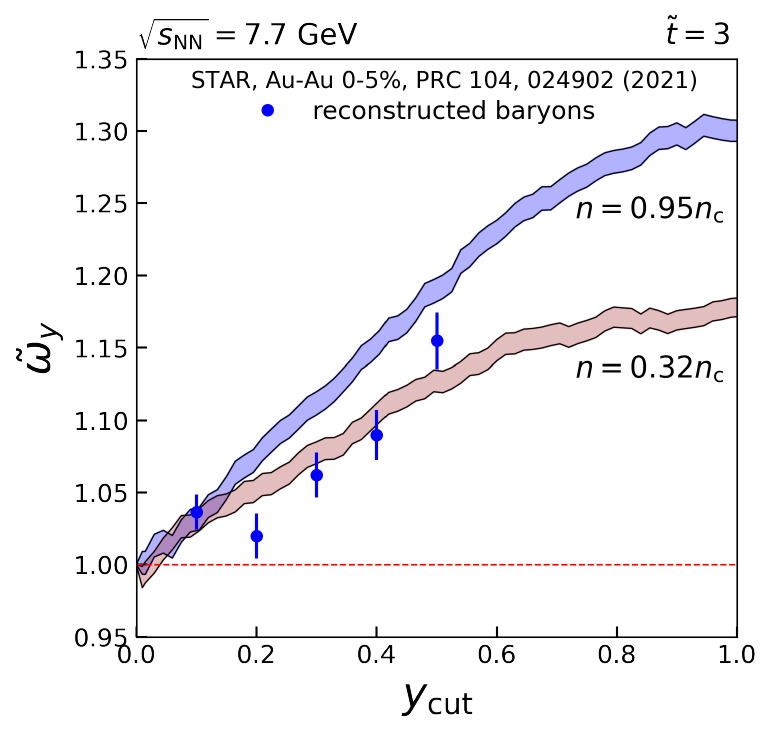}
    \caption{Corrected scaled variance as a function of rapidity cut $y_{\rm cut}$ for at fixed collision energy of $\sqrt{s_{\rm NN}}=7.7$ GeV, calculated with molecular dynamics at time $\tilde t = 3$ and density $n = 0.32 n_c$~(purple band) and $n = 0.95 n_c$~(blue band). 
    The blue points depict the processed experimental data~\cite{STAR:2021iop} of the STAR Collaboration for baryons and have the same meaning as in Fig.~\ref{fig-wsNN}.
    }
    \label{fig:wytycutdep}
\end{figure}

We, therefore, recalculate the behavior of fluctuations at different collision energies by analyzing the molecular dynamics data for $\tilde t = 3$, which is representative of the time scales relevant for heavy-ion collisions.
The corresponding results are shown in the right panel of Fig.~\ref{fig-wsNN}.
The enhancement of fluctuations for $n = 0.32 n_c$ and $n = 0.95 n_c$ is significantly suppressed by the finite-time effects, especially for $n = 0.95 n_c$.
We obtain that both $n = 0.32 n_c$ and $n = 0.95 n_c$ are in fair agreement with experimental data on $\tilde \omega_B$ at $\sNN = 7.7$~GeV.
This observation confirms that the data at $\sNN = 7.7$~GeV are consistent with the presence of sizable attractive interactions that enhance the scaled variance, although it does not pinpoint how close the system is to the CP.
Both the freeze-out of fluctuations near the CP~($n = 0.95 n_c$ case) or at a density considerably below the critical one~($n = 0.32 n_c$) are consistent with the data for $y_{\rm cut} = 0.5$.

Finite-time effects have a very mild effect on the calculation at $n = 1.9 n_c$ given that the time $\tilde t = 3$ is considerably larger than the corresponding equilibration time for fluctuations, which for $n = 1.9 n_c$ is $\tilde \tau_{\rm eq} \sim 1$~(Table~\ref{tab:taufits}).
Qualitatively, one can draw a conclusion that a system dominated by repulsive interactions shows a suppression of fluctuations, which locally equilibrate on faster time scales than those driven by attractive interactions.

\subsection{Acceptance dependence}

Figure~\ref{fig:wytycutdep} shows the dependence of $\tilde \omega$ at $\tilde t = 3$ on the value of rapidity cut $y_{\rm cut}$ at the collision energy $\sNN = 7.7$~GeV, along with the available experimental data.
The calculations indicate that $\tilde \omega$ continues to increase with $y_{\rm cut}$ unit the acceptance covers two units of rapidity~($y_{\rm cut} = 1$).
$\tilde \omega$ has a maximum around $y_{\rm cut} \simeq 1$ and the decreases with $y_{\rm cut}$ at larger acceptances.
This behavior can be mapped to the symmetric shape of $\tilde \omega$ with respect to $\alpha_y$ shown in Fig.~\ref{fig:wyalpha}, with $y_{\rm cut} = 1$ approximately corresponding to $\alpha_y = 0.5$.
We note that the behavior of $\tilde \omega$ at $\alpha_y > 0.5$ maybe be sensitive to the choice of (periodic) boundary conditions and thus may not directly apply to heavy-ion collisions.
For this reason we only show the results up to $y_{\rm cut} = 1$.

The experimental data are available up to $y_{\rm cut} = 0.5$, shown in Fig.~\ref{fig:wytycutdep} for the reconstructed baryons.
The data agrees qualitatively with both the $n = 0.32 n_c$ and $n = 0.95 n_c$ cases, indicating that it may be challenging to distinguish whether the system is close to the CP or not.
Quantitatively, a better agreement is seen for the $n = 0.32 n_c$ case, however, we must emphasize again here that our model is not yet suited for drawing robust quantitative conclusions.
Both cases predict continued growth up to $y_{\rm cut} = 1$, which can be verified with BES-II measurements utilizing expanded rapidity coverage.

\subsection{Limitations}\label{Limit}

It is important to emphasize the limitations of our approach when applied to measurements from heavy-ion collisions.
The main caveat is that we do not simulate the full dynamics of heavy-ion collisions but perform box simulations of the subsystem of nucleons near the CP, and implement the collective flow effect on top of these simulations.
It is thus assumed that our LJ molecular dynamics simulation models the behavior of (critical) fluctuations nucleons in the local rest frame, where we rely heavily on the universality of critical behavior,
while collective expansion is described through a separate mechanism.
In particular, this implies that, in our approach, particles in a box reach equilibrium\footnote{In the case of a finite-time calculation~($\tilde \tau = 3$), the equilibrium may be incomplete.} first and then the shift of $y^{\rm acc}$ is applied.
On the other hand, local equilibrium may be maintained at best in heavy-ion collisions, while the system never reaches a global equilibrium due to the continued expansion.
Furthermore, we also neglect event-by-event fluctuations of the longitudinal flow, as well as fluctuations and the inhomogeneous rapidity distribution of baryon stopping.

Our simulations are performed at fixed values of temperature and nucleon number densities, corresponding to an idealized picture of fluctuations being determined at a fixed point on the phase diagram at each collision energy, for instance, at chemical freeze-out.
As heavy-ion collisions are highly dynamic processes, the fluctuations may instead reflect the history of the collision, which is characterized by different temperatures and densities.
Furthermore, even at freeze-out, the densities and temperatures are different at different collision energies, therefore, the bands shown in Fig.~\ref{fig-wsNN} should not be viewed as predictions for the possible $\sNN$-dependence of the scaled variances. Rather, these calculations indicate the expectation for the possible value of $\tilde \omega_\alpha$ at a given $\sNN$ if collisions at this energy correspond to a certain point on the phase diagram relative to the CP location.
Furthermore, our simulations are performed for a non-relativistic system without incorporating any mesonic or partonic degrees of freedom.

Due to the above limitations, our results should mainly be viewed as qualitative expectations for the possible CP signals in fluctuations, which nevertheless do include such essential effects as exact baryon conservation, difference between coordinate and momentum space, and finite-size and finite-time effects.

Our implementation can be improved in different ways to make the predictions more quantitative.
For instance, instead of a single box, we could consider a collection of boosted fireballs along the longitudinal axis, each described by a separate LJ system in the local rest frame.
This could mimick better the local equilibrium of the expanding system in heavy-ion collisions but will have to be accompanied by an analysis of the meaningful box size, conservation laws,
cross-talk between the boxes, and so on.
Another possibility would be incorporating the interactions responsible for critical fluctuation dynamics into transport model framework such as SMASH~\cite{SMASH:2016zqf} or UrQMD~\cite{Bass:1998ca,Bleicher:1999xi}.

\section{Discussion and summary}
\label{sec-discussion}
In this paper, we studied the behavior of fluctuations, namely the scaled variance, near the critical point by performing molecular dynamics simulations of the Lennard-Jones fluid. 
The simulations were performed in a box with periodic boundary conditions along the supercritical isotherm $T = 1.06 T_c$, where $T_c$ is the critical point temperature. 
As a microscopic model calculation, it naturally incorporated effects such as correlation length, exact conservation laws, and finite system size.
Compared to our previous work~\cite{Kuznietsov:2022pcn}, we have incorporated additional phenomena, such as ensemble averaging, longitudinal collective expansion, and finite-time effects, to bring our calculations closer to the conditions encountered relativistic heavy-ion collisions.
We summarize our main findings as follows:
\begin{itemize}
    \item We observe large particle number fluctuations in coordinate space near the critical point of the Lennard-Jones system when calculating them as ensemble averages. The results are in quantitative agreement~(Figs.~\ref{fig-ens-average1} and \ref{fig-wcoord}) with our earlier study~\cite{Kuznietsov:2022pcn} that employed time averaging, confirming that the ergodic hypothesis holds for fluctuations. In the context of heavy-ion collisions, if one interprets these events as samples from an ensemble, as is commonly done, this observation confirms the suitability of fluctuations for the search for critical behavior.

    \item Analysis of the time dependence allowed us to elucidate equilibration dynamics of fluctuations. Generally, the equilibration time $\tilde \tau_\alpha$ depends on the choice of acceptance in which the fluctuations are analyzed. By comparing the values of $\tilde \tau_\alpha$ for the same acceptance but different densities, we observe indications for a (local) maximum in the dependence of equilibration time on density in the vicinity of the critical density~(Table~\ref{tab:taufits}), meaning that fluctuations near the CP take more time to develop. 

    \item The presence of collective flow is crucial for observing large fluctuations in momentum space acceptance relevant for experimental measurements. For sufficiently strong collective flow, such as the Bjorken flow at high energies, the momentum~(rapidity) space fluctuations reflect those in coordinate space~(Figs.~\ref{fig:wybeam} and~\ref{fig:wyalpha}).

    \item Fluctuations near the critical point measured in acceptance spanning one unit at midrapidity, $|y| < 0.5$, show the maximum value in collision energy range $\sNN \sim 5-7$~GeV, indicating that these collision energies are optimal for the search of critical behavior is it exists in a heavy-ion regime.
    It may be counterintuitive that the strongest signal is observed at the intermediate collision energies rather than at the highest collision energies where the longitudinal flow is the strongest.    
    This comes from an interplay between flow and finite system-size effects: the increase of the signal with $\sNN$ due to stronger flow is compensated by larger finite-size effects, given that a fixed value of rapidity cut corresponds to a smaller number of particles~(baryons) inside the acceptance at higher $\sNN$, effectively corresponding to a smaller size of the system~(smaller number of particles) captured inside the acceptance.

    \item Experimental data of the STAR Collaboration on proton number scaled variance shows enhancement of fluctuations at lowest BES-I energies relative to the baseline of unity when the $1-\alpha$ correction for baryon conservation is accounted for. 
    In particular, we find $\tilde \omega_p \simeq 1.06$ for protons and $\tilde \omega_B \simeq 1.15$ for baryons at $\sNN = 7.7$~GeV.
    Interestingly, these values agree with the corresponding molecular dynamics calculation near the CP~($T = 1.06 T_c, \, n = 0.95 n_c$), incorporating finite-size and finite-time effects.
    Therefore, the experimental data at $\sNN = 7.7$~GeV is compatible with the freeze-out of fluctuations near the CP, although this does not rule out other scenarios.
    Our results do motivate a detailed analysis of proton number cumulants in collision energy range $\sNN \sim 3-10$~GeV, which will be filled with experimental measurements from RHIC-BES-II, RHIC-FXT, and CBM-FAIR programs in the foreseeable future.

\end{itemize}

It should be emphasized that our present approach needs further improvements for more quantitative applications to experimental measurements.
In particular, the system in our simulations evolves at a constant particle number and energy density, which is not the case for heavy-ion collisions. 
As mentioned before, we also neglect the production of antibaryons, which would be required for applications at $\sNN \gtrsim 20$~GeV.
The modeling can also be improved by considering more realistic longitudinal density and flow profiles~(as opposed to the Bjorken-like picture employed here), their event-by-event fluctuations, as well as incorporating transverse expansion and $p_T$ cuts. 
One can also implement local equilibrium relevant for heavy-ion collisions by considering a collection of boosted fireballs along the longitudinal axis instead of a single fireball.
Each fireball would be described by a separate LJ system in the local rest frame and will have to be accompanied by an analysis of the meaningful box size, conservation laws,
cross-talk between the boxes etc.
These extensions will be the subject of future studies.

We also plan to explore the behavior of high-order~(non-Gaussian) cumulants, such as skewness and kurtosis. 
On the one hand, these are expected to exhibit increased sensitivity to the CP. 
On the other hand, high-order cumulants may also be affected by finite-size and finite-time effects.
Studying the cumulants of different order within a single microscopic description will allow us to elucidate which observables are most promising in the search for critical behavior.

Another potential avenue is the study of the mixed-phase region of the first-order phase transition and its possible signatures in expanding systems created in heavy-ion collisions. In particular, the production of clusters~(light nuclei) can be particularly sensitive to the existence of mixed-phase and the associated critical point.

\begin{acknowledgments}

\emph{Acknowledgments.} 
We thank Carsten Greiner, Roman Poberezhnyuk, Claudia Ratti and Yuri Sinyukov for discussions.
The authors acknowledge the use of the PhysGPU Cluster and the support from the Research Computing Data Core at the University of Houston to carry out the research presented here.
V.K. has been supported by the U.S. Department of Energy, 
Office of Science, Office of Nuclear Physics, under contract number 
DE-AC02-05CH11231.
\end{acknowledgments}

\bibliography{main}

\end{document}